\documentclass[doublecol,amsmath,amssymb,revision]{epl2}
\usepackage{amssymb}
\usepackage{graphicx}
\usepackage{dcolumn}
\usepackage{bm}
\usepackage{hhline}
\usepackage{longtable}
\usepackage[usenames]{color}

\begin{document}

\title{Enumeration of spanning trees in a pseudofractal scale-free web}
\shorttitle{Enumeration of spanning trees in a pseudofractal
scale-free web}

\author{Zhongzhi Zhang\inst{1,2} \footnote{ \email{zhangzz@fudan.edu.cn} } \and Hongxiao Liu \inst{1,2} \and Bin Wu\inst{1,2}  \and  Shuigeng Zhou\inst{1,2} \footnote{\email{sgzhou@fudan.edu.cn}}}
\shortauthor{Zhongzhi Zhang, Hongxiao Liu, Bin Wu, Shuigeng Zhou}

 \institute{
  \inst{1} School of Computer Science, Fudan University, Shanghai 200433, China\\
  \inst{2} Shanghai Key Lab of Intelligent Information Processing, Fudan University, Shanghai 200433, China}

\date{\today}

\begin{abstract}{
Spanning trees are an important quantity characterizing the
reliability of a network, however, explicitly determining the number
of spanning trees in networks is a theoretical challenge. In this
paper, we study the number of spanning trees in a small-world
scale-free network and obtain the exact expressions. We find that
the entropy of spanning trees in the studied network is less than 1,
which is in sharp contrast to previous result for the regular
lattice with
the same average degree, the entropy of which is higher than 1. 
Thus, the number of spanning trees in the scale-free network is much
less than that of the corresponding regular lattice. We present that
this difference lies in disparate structure of the two networks. 
Since scale-free networks are more robust than regular networks
under random attack, our result can lead to the counterintuitive
conclusion that a network with more spanning trees may be relatively
unreliable.}
\end{abstract}



 \maketitle

\section{Introduction}

The enumeration of spanning trees in networks (graphs) is a
fundamental issue in mathematics~\cite{BuPe93,Ly05,LyPeSc08},
physics~\cite{Wu77,ChChYa07}, and other discipline~\cite{JaTh89}. A
spanning tree of any connected network is defined as a minimal set
of edges that connect every node. The problem of spanning trees is
relevant to various aspects of networks, such as
reliability~\cite{Bo86,SzAlKe03}, optimal
synchronization~\cite{NiMo06}, standard random walks~\cite{NoRi04},
and loop-erased random walks~\cite{DhDh97}. In particular, the
number of spanning trees corresponds to the partition function of
the $q$-state Potts model~\cite{Wu82} in the limit of $q$
approaching zero, which in turn closely relates to the sandpile
model~\cite{BaTaWi87}.

Because of the diverse applications in a number of
fields~\cite{WuCh04}, a lot of efforts have been devoted to the
study of spanning trees. For example, the exact number of spanning
trees in regular lattices~\cite{Wu77,ShWu00} and Sierpinski
gaskets~\cite{ChChYa07} has been explicitly determined in previous
studies. However, regular lattices and fractals cannot well mimic
the real-life networks, which have been recently found to
synchronously exhibit two striking properties: scale-free
behavior~\cite{BaAl99} and small-world effects~\cite{WaSt98} that
has a strong impact on the enumeration problems on networks. For
example, previous work on counting subgraphs, such as
cliques~\cite{BiMa06}, loops and Hamiltonian cycles~\cite{BiMa05},
has shown that scale-free degree distribution implies a very
non-trivial structure of subgraphs. However, so far investigation on
the number of spanning trees in scale-free small-world networks has
been still missing. In view of the distinct structure, as compared
to regular lattices, it is of great interest to examine spanning
trees in scale-free small-world networks.

In this paper, we intend to fill this gap by providing a first
analytical research of spanning trees in a small-world network with
inhomogeneous connectivity. In order to exactly obtain the number of
spanning trees, by using a renormalization group
method~\cite{KnVa86}, we consider a deterministically growing
scale-free network with small-world effect. We find that the entropy
of its spanning trees is smaller than 1, which is a striking result
that is qualitatively different from that of two dimensional regular
lattices with identical average degree, in which the entropy is
higher than 1. Thus, the number of its spanning trees is much lower
than that of its corresponding regular lattice.
We show that this difference can be accounted for by the
heterogeneous structure of scale-free networks. Since the network
under study is much robust to random deletion of edges, as opposed
to regular lattice, our result suggests that networks with more
spanning trees are not always more stable to random breakdown of
edges, compared with those networks with less spanning trees.

\begin{figure}
\begin{center}
\includegraphics[width=.70\linewidth,trim=60 50 60 40]{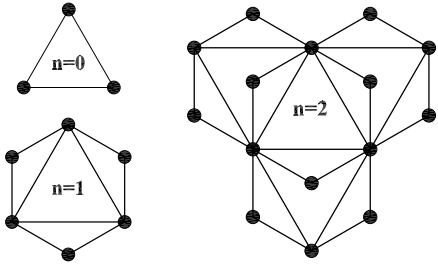} \
\end{center}
\caption[kurzform]{The first three generations of the iterative
scale-free network.} \label{network}
\end{figure}

\section{Pseudofractal scale-free web}

\begin{figure}
\begin{center}
\includegraphics[width=.60\linewidth,trim=40 25 40 15]{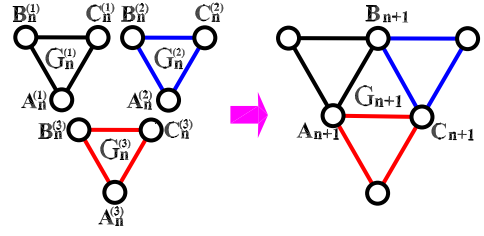}
\caption{(Color online) Second construction method of the network.
$G_{n+1}$ can be obtained by joining three copies of $G_{n}$ denoted
as $G_{n}^{(\eta)}$ $(\eta=1,2,3)$, the three hubs of which are
represented by $A_{n}^{(\eta)}$, $B_{n}^{(\eta)}$, and
$C_{n}^{(\eta)}$. In the merging process, hubs $A_{n}^{(1)}$ (resp.
$C_{n}^{(1)}$, $A_{n}^{(2)}$) and $B_{n}^{(3)}$ (resp.
$B_{n}^{(2)}$, $C_{n}^{(3)}$) are identified as a hub node $A_{n+1}$
(resp. $B_{n+1}$, $C_{n+1}$) in $G_{n+1}$.} \label{merge}
\end{center}
\end{figure}

The studied scale-free network~\cite{DoGoMe02,ZhQiZhXiGu09}, denoted
by $G_n$ after $n$ ($n\geq 0$) generations, is constructed as
follows: For $n=0$, $G_0$ is a triangle. For $n\geq 1$, $G_{n}$ is
obtained from $G_{n-1}$: every existing edge in $G_{n-1}$ introduces
a new node connected to both ends of the edge. Figure~\ref{network}
illustrates the construction process for the first three
generations. The network exhibits some typical properties of real
networks. Its degree distribution $P(k)$ obeys a power law $P(k)\sim
k^{1+\ln 3/\ln 2}$, the average distance scales logarithmically with
network order (number of nodes)~\cite{ZhZhCh07}, and the clustering
coefficient is $\frac{4}{5}$. Alternatively, the network can be also
created in another method~\cite{ZhZhCh07,Bobe05}. Given the
generation $n$, $G_{n+1}$ may be obtained by joining at the hubs
(the most connected nodes) three copies of $G_{n}$, see
Fig.~\ref{merge}. According to the latter construction algorithm, we
can easily compute the network order of $G_n$ is $V_n=\frac{3^{n+1}
+3}{2}$. In $G_n$, there are three hubs denoted by $A_{n}$, $B_{n}$,
and $C_{n}$, respectively.

\section{Number of spanning trees}

After introducing the network construction and its properties, next
we will study both numerically and analytically spanning trees in
this scale-free network.

\subsection{Numerical solution}

\begin{figure}
\begin{center}
\includegraphics[width=0.8\linewidth,trim=60 45 60 40]{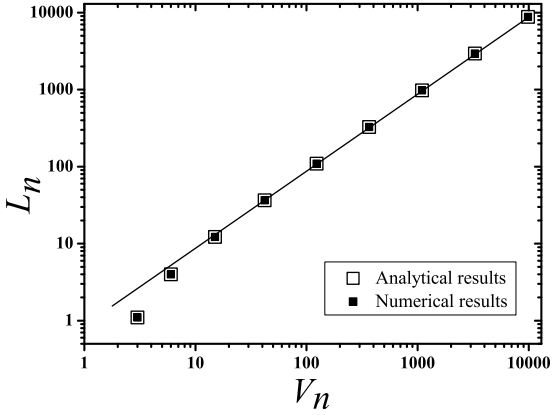}
\end{center}
\caption[kurzform]{\label{SpanTree01} Logarithm of the number of
spanning trees $N_{\rm ST}(n)$ in network $G_n$ as a function of
network order $V_n$ on a log-log scale. In the figure, $L_n=\ln
N_{\rm ST}(n)$; the filled symbols are the numerical results
obtained from Eq.~(\ref{ST01}), while the empty symbols correspond
to the exact values from Eq.~(\ref{ST04}), both of which completely
agree with each other.}
\end{figure}

According to the well-known result~\cite{Bi93}, we can obtain
numerically but exactly the number of spanning trees, $N_{\rm
ST}(n)$, by computing the non-zero eigenvalues of the Laplacian
matrix  associated with $G_{n}$ as
\begin{equation}\label{ST01}
 N_{\rm ST}(n)=\frac{1}{V_n}\prod_{i=1}^{i=V_n-1}\lambda_i(n)\,,
\end{equation}
where $\lambda_i(n)$ ($i = 1, 2,\ldots, V_n-1$) are the $V_n-1$
nonzero eigenvalues of the Laplacian matrix for $G_{n}$. For a
network, the non-diagonal element $l_{ij}$ ($i \neq j$) of its
Laplacian matrix is -1 (or 0) if nodes $i$ and $j$ are (or not)
directly connected, while the diagonal entry $l_{ii}$ equals the
degree of node $i$. Using Eq.~(\ref{ST01}), we can calculate
directly the number of spanning trees $N_{\rm ST}(n)$ of $G_{n}$
(see Fig.~\ref{SpanTree01}). From Fig.~\ref{SpanTree01}, we can see
that $N_{\rm ST}(n)$ approximately grows exponentially in $V_n$.
This allows to define the entropy of spanning trees for $G_{n}$ as
the limiting value~\cite{BuPe93,Ly05,LyPeSc08}
\begin{equation}\label{Entropy}
E_{G_n}=\lim_{V_n \rightarrow \infty }\frac{\ln N_{\rm
ST}(n)}{V_n}\,,
\end{equation}
which is a finite number and a very interesting quantity
characterizing the network structure.

It should be mentioned that although the expression of
Eq.~(\ref{ST01}) seems compact, the computation of eigenvalues of a
matrix of order $V_{n} \times V_{n}$ makes heavy demands on time and
computational resources for large networks. Thus, one can count the
number of spanning trees by directly calculating the eigenvalues
only for the first several iterations, which is not acceptable for
large graphs. Particularly, by using the eigenvalue method it is
difficult and even impossible to obtain the entropy $E_{G_n}$. It is
thus of significant practical importance to develop a
computationally cheaper method for enumerating spanning trees that
is devoid of calculating eigenvalues. Fortunately, the iterative
network construction permits to calculate recursively $N_{\rm
ST}(n)$ and $E_{G_n}$ to obtain exact solutions.

\subsection{Closed-form formula}

\begin{figure}
\begin{center}
\includegraphics[width=0.9\linewidth,trim=0 100 0 50]{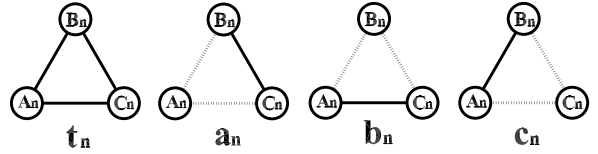}
\end{center}
\caption[kurzform]{\label{Demo01} Illustrative definition for the
spanning subgraphs of $G_n$. The two hub nodes connected by a solid
line are in one tree, and the two hub nodes linked by a dotted line
belong to different trees.}
\end{figure}

To get around the difficulties of the eigenvalue method, we use an
analytic technique based on a decimation procedure~\cite{KnVa86}.
For simplicity, we use $t_n$ to express $N_{\rm ST}(n)$. Moreover,
let $a_n$ denote the number of spanning subgraphs of $G_n$
consisting of two trees such that the hub node $A_n$ belongs to one
tree and the two other hubs ($B_n$ and $C_n$) are in the other tree.
Analogously, we can define quantities $b_n$ and $c_n$, see
Fig.~\ref{Demo01}. By symmetry, we have $a_n=b_n=c_n$. Thus, in the
following computation, we will replace $b_n$ and $c_n$ by $a_n$.

\begin{figure}
\begin{center}
\includegraphics[width=1.00\linewidth,trim=0 60 0 30]{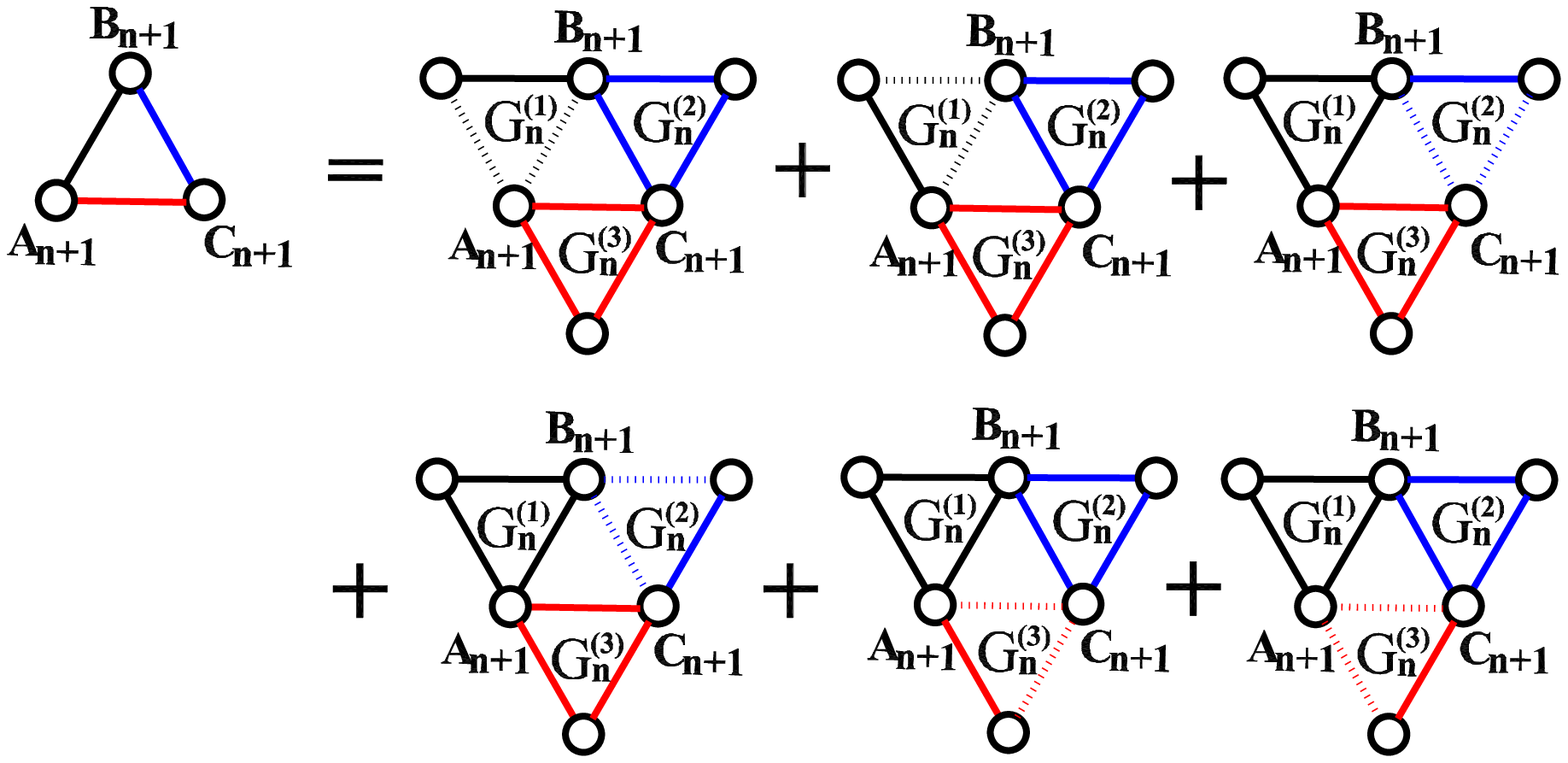}
\end{center}
\caption[kurzform]{\label{Demo02} (Color online) Illustration for
the recursion expression for the number of spanning trees $t_{n+1}$
in $G_{n+1}$. The two nodes at both ends of a solid line are in one
tree, while the two nodes at both ends of a dotted line are in
separate trees.}
\end{figure}

\begin{figure}
\begin{center}
\includegraphics[width=0.90\linewidth,trim=50 20 50 0]{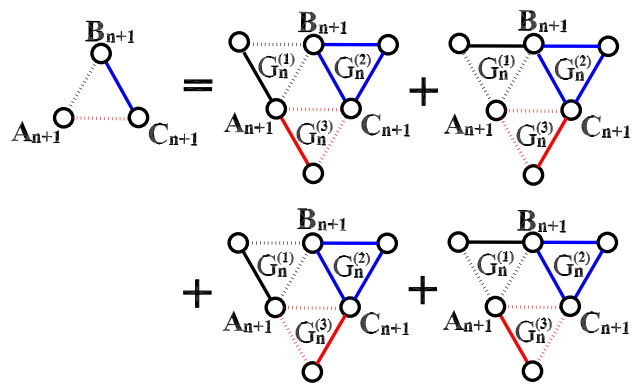}
\end{center}
\caption[kurzform]{\label{Demo03} (Color online) Illustration for
the recursive expression for the number of spanning subgraphs
$a_{n+1}$ corresponding to network $G_{n+1}$. The two nodes at both
ends of a solid line (dotted line) are in one tree (two trees).}
\end{figure}

Considering the self-similar network structure, the following
fundamental relations can be established:
\begin{equation}\label{ST02}
t_{n+1}=(t_{n})^2(a_n+c_n+a_n+b_n+c_n+b_n)=6a_n(t_{n})^2\,
\end{equation}
and
\begin{equation}\label{SG01}
a_{n+1}=t_{n}[(c_n)^2+a_nb_n+b_nc_n+a_nc_n]=4t_{n}(a_n)^2\,.
\end{equation}
Equation~(\ref{ST02}) can be explained as follows. Since $G_{n+1}$
is obtained via merging three $G_{n}$ by identifying three couples
of hub nodes, to get the number of spanning trees $t_{n+1}$ for
$G_{n+1}$, one of the copies of $G_{n}$ must be spanned by two
trees. There are six possibilities as shown in Fig.~\ref{Demo02},
from which it is easy to derive Eq.~(\ref{ST02}). Analogously,
Eq.~(\ref{SG01}) can be understood based on Fig.~\ref{Demo03}.

To obtain $t_n$, we define an intermediary variable
$h_n=\frac{t_n}{a_n}$ that obeys the following recursive relation
\begin{equation}\label{Mid01}
h_{n+1}=\frac{t_{n+1}}{a_{n+1}}=\frac{3t_{n}}{2a_{n}}=\frac{3}{2}h_{n}\,.
\end{equation}
With the initial condition $t_0=3$ and $a_0=1$, we have $h_0=3$.
Hence, Eq.~(\ref{Mid01}) is solved to yield
\begin{equation}\label{Mid02}
h_{n}=\frac{3^{n+1}}{2^{n}}\,.
\end{equation}
Then,
\begin{equation}\label{Mid03}
a_{n}=\frac{2^{n}}{3^{n+1}}t_{n}\,.
\end{equation}
Plugging this expression into Eq.~(\ref{ST02}) leads to
\begin{equation}\label{Mid04}
t_{n+1}=\frac{2^{n+1}}{3^{n}}(t_{n})^{3}\,.
\end{equation}
Considering the initial value $t_0=3$, we can solve
Eq.~(\ref{Mid04}) to obtain the explicit solution
\begin{eqnarray}\label{ST03}
N_{\rm ST}(n)=t_n=2^{({3^{n + 1}}-2n-3)/4}3^{({3^{n+1}}+2n+1)/4}\,.
\end{eqnarray}
Analogously, we can derive the exact formula for $a_{n}$ as:
\begin{eqnarray}\label{Mid05}
a_n=2^{({3^{n+1}}+2n-3)/4}3^{({3^{n+1}}-2n-3)/4}\,.
\end{eqnarray}

It not difficult to represent $N_{\rm ST}(n)$ as a function of the
network order $V_n$, with the aim to obtain the relation between the
two quantities. Recalling $V_n=\frac{3^{n+1} +3}{2}$, we have
$3^{n+1}=2V_n-3$ and $n+1=\ln(2V_n-3)/\ln3$. These relations enable
one to write $N_{\rm ST}(n)$ in terms of $V_n$ as
\begin{eqnarray}\label{ST04}
N_{\rm ST}(n)= 2^{[V_n-\ln(2V_n-3)/\ln
3-2]/2}3^{[V_n+\ln(2V_n-3)/\ln 3-2]/2}\,.
\end{eqnarray}

We have confirmed the closed-form expressions for $N_{\rm ST}(n)$
against direct computation from Eq.~(\ref{ST01}). In the full range
of $0\leq n \leq 8$, they are perfectly consistent with each other,
which shows that the analytical formulas provided by
Eqs.~(\ref{ST03}) and~(\ref{ST04}) are right.
Figure~\ref{SpanTree01} shows the comparison between the numerical
and analytical results.

Equation~(\ref{ST04}) unveils the explicit dependence relation of
$N_{\rm ST}(n)$ on the network order $V_n$. Inserting
Eq.~(\ref{ST04}) into Eq.~(\ref{Entropy}), it is easy to obtain the
entropy of spanning trees for $G_n$ given by
\begin{equation}\label{Entropy02}
E_{G_n}=\lim_{V_n \rightarrow \infty }\frac{\ln N_{\rm
ST}(n)}{V_n}=\frac{1}{2}(\ln2+\ln 3)\simeq 0.89588\,.
\end{equation}
This obtained asymptotic value is the smallest entropy (lower than
1) that has not been reported earlier for other networks with an
average degree of 4. For example, the entropy for spanning trees in
the square lattice is 1.16624~\cite{Wu77}, a value larger than 1.
Thus, the number of spanning trees in $G_n$ is much less than that
in the square lattice with the same average degree of nodes.


From the result obtained above, we can conclude that the
pseudofractal scale-free network has less spanning trees than the
regular lattice with the same average degree. The difference can be
attributed to the structural characteristics of the two classes of
networks. In scale-free networks, nodes have a heterogeneous
connectivity, which leads to an inhomogeneous distribution of
Laplacian spectra~\cite{DoGoMe02,ChLuVu03,ZhChYe10}. On the
contrary, in regular lattices, since all nodes have approximately
the same degree, their Laplacian spectra have a homogenous
distribution. Thus, for two given scale-free and regular networks
with the same order and average node degree, the sum of the
eigenvalues of their Laplacian matrices are the same, but the
product of non-zero Laplacian spectra of the scale-free network is
smaller than its counterpart of the regular network, because of the
different distributions of the Laplacian spectra resulting from
their distinct connectivity distribution. Hence, the heterogeneous
structure is responsible for the difference of number of spanning
trees in scale-free networks and regular lattices. It should be
stressed that although we only study a specific deterministic
scale-free network, we expect to find a qualitatively similar result
about spanning trees in real-world scale-free networks, since they
have similar structural characteristics as that discussed above.

As an important invariant of a network, the number of spanning trees
is a relevant measure of the reliability of the network.
Intuitively, among all connected graphs with the same numbers of
nodes and edges, networks having more spanning trees are more
resilient (reliable) to the random removal of edges, compared with
those with less spanning trees. That is to say, the former has a
larger threshold of bond percolation than that of the latter.
However, recent work~\cite{AlJeBa00,CaNeStWa00,CoErAvHa01} have
shown that inhomogeneous networks, such as scale-free networks, are
impressively robust than homogeneous networks (e.g., exponential
networks and regular networks) with respect to random deletion of
edges. Thus, combining with our above result, we can reach the
following counterintuitive conclusion that networks (e.g.,
scale-free networks) with less spanning trees do not mean more
vulnerable to random breakdown of links than those (e.g., regular
lattices) with more spanning trees.

\section{Conclusions}

In summary, diverse real-life networks possess power-law degree
distribution and small-world effect. In this paper, we have studied
and enumerated explicitly the number of spanning trees in a
scale-free network with small-world behavior. The exact solution was
obtained on the basis of some precise recursion relations derived
from the iterative construction of the network addressed. It was
demonstrated that scale-free network has much less spanning trees
compared to the regular lattice with the same number of nodes and
edges. It was shown that this difference is rooted in the inherent
architecture of the two types of networks. Although it is generally
thought that increasing the number of spanning trees over all
networks with identical number of nodes and edges can lead to a less
fragile network, our results strikingly indicate otherwise. Our work
may be helpful for designing and improving the reliability of
networks.

\section{Acknowledgment}

The authors would like to thank Yichao Zhang and Yuan Lin for their
assistance. This research was supported by the National Natural
Science Foundation of China under Grants No. 60704044 and No.
60873070, the National Basic Research Program of China under Grant
No. 2007CB310806, and Shanghai Leading Academic Discipline Project
No. B114.

\end{document}